\pdfoutput=1

 \documentclass[final,3p,times,twocolumn]{elsarticle}

\usepackage{amsfonts,amssymb,amsmath,graphicx,multirow,hyperref}

\newcommand{\degree}{\ifmmode {^{\circ}} \else {$^{\circ}$} \fi}
\newcommand{\degrees}{\ifmmode {^{\circ}} \else {$^{\circ}$} \fi}

\newcommand{\unit}[1]{\ifmmode {\rm\ #1\,} \else {$\rm #1$} \fi}
\newcommand{\quarter}{\ifmmode {\frac{1}{4}} \else {$\frac{1}{4}$} \fi}

\newcommand{\etal}{{\em et al.~}}

\newcommand{\s}{\unit{s}}
\newcommand{\us}{\unit{\mu s}}

\newcommand{\MHz}{\unit{MHz}}
\newcommand{\GHz}{\unit{GHz}}

\newcommand{\tten}[1]{\ifmmode {\times 10^{#1}} \else {$\times 10^{#1}$} \fi}
\newcommand{\tentothe}[1]{\ifmmode {10^{#1}} \else {$10^{#1}$} \fi}

\newcommand{\dm}{\unit{pc\ cm^{-3}}}

\newcommand{\doublet}{\ifmmode {\lambda\lambda} \else {$\lambda\lambda$} \fi}
\newcommand{\singlet}{\ifmmode {\lambda} \else {$\lambda$} \fi}

\begin{document}

\begin{frontmatter}



\title{New SETI Sky Surveys for Radio Pulses}

\author[label1,label5]{Andrew Siemion}
\author[label3]{Joshua Von Korff}
\author[label5,label6]{Peter McMahon}
\author[label2]{Eric Korpela}
\author[label2,label5]{Dan Werthimer}
\author[label2]{David Anderson}
\author[label1]{Geoff Bower}
\author[label2]{Jeff Cobb}
\author[label1]{Griffin Foster}
\author[label2]{Matt Lebofsky}
\author[label1]{Joeri van Leeuwen}
\author[label5]{Mark Wagner}

\address[label1]{University of California, Berkeley - Department of Astronomy, Berkeley, California, USA}
\address[label2]{University of California Berkeley - Space Sciences Laboratory, Berkeley, California, USA}
\address[label3]{University of California, Berkeley - Department of Physics, Berkeley, California, USA}
\address[label5]{University of California, Berkeley - Berkeley Wireless Research Center, Berkeley, California, USA}
\address[label6]{Stanford University - Department of Computer Science, Stanford, California, USA}


\begin{abstract}
Berkeley conducts 7 SETI programs at IR, visible and radio wavelengths.  Here we review two of the newest efforts, Astropulse and Fly's Eye.

A variety of possible sources of microsecond to millisecond radio pulses have been suggested in the last several decades, among them such exotic events as evaporating primordial black holes, hyper-flares from neutron stars, emissions from cosmic strings or perhaps extraterrestrial civilizations, but to-date few searches have been conducted capable of detecting them.  The recent announcement by Lorimer et al. of the detection of a powerful ($\approx 30$ Jy) and highly dispersed ($\approx 375$ cm$^{-3}$ pc) radio pulse in Parkes multi-beam survey data has fueled additional interest in such phenomena. 

We are carrying out two searches in hopes of finding and characterizing these $\mu$s to ms time scale dispersed radio pulses.  These two observing programs are orthogonal in search space; the Allen Telescope Array's (ATA) ``Fly's Eye" experiment observes a 100 square degree field by pointing each 6m ATA antenna in a different direction; by contrast, the Astropulse sky survey at Arecibo is extremely sensitive but has 1/3,000 of the instantaneous sky coverage.  Astropulse's multibeam data is transferred via the internet to the computers of millions of volunteers.  These computers perform a coherent de-dispersion analysis faster than the fastest available supercomputers and allow us to resolve pulses as short as 400 ns.  Overall, the Astropulse survey will be 30 times more sensitive than the best previous searches.  Analysis of results from Astropulse is at a very early stage.  

The Fly's Eye was successfully installed at the ATA in December of 2007, and to-date 
approximately 450 hours of observation has been performed.  We have detected three pulsars 
(B0329+54, B0355+54, B0950+08) and six giant pulses from the Crab pulsar in our diagnostic 
pointing data.  We have not yet detected any other convincing bursts of astronomical origin in our 
survey data. 
\end{abstract}




\end{frontmatter}


\section{Introduction}
The Berkeley SETI group conducts seven searches at visible, IR and radio 
wavelengths, covering a wide variety of signal
types and spanning a large range of time scales:
SETI@home searches for radio signals with time scales ranging
from mS to seconds.
SEVENDIP searches for nS time scale pulses at visible wavelengths.
ASTROPULSE and Fly's Eye search for dispersed uS and mS time scale radio 
pulses from  extraterrestrial civilizations, pulsars, or evaporating 
primordial black holes.  SERENDIP and SPOCK search for continuous narrow 
band  signals in the radio and optical bands respectively.  DYSON 
searches for infrared excess from advanced civilizations that use a lot 
of energy.

SETI@home II, ASTROPULSE and SERENDIP V are radio sky surveys at the
300 meter Arecibo telescope.
Commensal observations have been conducted almost continuously
for the past ten years and are ongoing.
Most beams on the sky visible to the Arecibo telescope have been
observed four or more times.
We rank SERENDIP and SETI@home candidate signals based on
the number of independent observations,
the strength of the signals,
the closeness of the signals in frequency and sky position,
and the proximity to stars, planetary systems,
galaxies, and other interesting astronomical objects.
SETI@home uses the CPU power of volunteered PCs to analyze data.
Five million people in 226 countries have participated.
Combined, their PCs form Earth's second most powerful supercomputer,
averaging 482 TeraFLOPs and contributing over two million years of CPU time.
Here we describe two of the newest SETI searches, Astropulse and Fly's Eye.

\section{SETI Pulse Searches}
One common assumption of SETI is that an alien civilization wishing to make 
contact with others would broadcast a signal that is easily detected
and easily distinguished from natural sources of radio emission.  One
way of achieving these goals is to send a narrow band signal.  By 
concentrating the signal power in a very narrow frequency band, the signal
can be made to stand out among the natural, broad-band sources of noise.
By the same token, signals leaked from a civilization's internal 
communications may be narrow band, but would be significantly
weaker than a direct attempt at extraterrestrial contact.

Because of this, radio SETI efforts have concentrated on detecting narrow
band signals.  When searching for narrow band signals it is best to use
a narrow search window (or channel) around a given frequency.  The wider the
channel, the more broadband noise is included in addition to any 
signal.  This broadband noise limits the sensitivity of the system.
Early systems used analog technology to create narrow 
bandpass filters that could observe at a single frequency channel.  
More recent systems use massive banks of dedicated 
fast Fourier transform (FFT)
processors to separate incoming signals into up to a billion channels,
each of width $\sim$1 Hz.

There are, however, limitations to this technique.  One limitation is that
extraterrestrial signals are unlikely to be stable in frequency due to 
accelerations of
the transmitter and receiver.  For example, a receiver listening for signals 
at 1.4 GHz located on the surface of the earth undergoes acceleration
of up to 3.4 cm/s$^2$ due to the earth's rotation.  This corresponds to 
a Doppler drift rate of 0.16 Hz/s.  If uncorrected,
an alien transmission would drift out of a 1 Hz channel in about 6 
seconds, effectively limiting the maximum integration time to 6 seconds.  
Because of the inverse relationship between maximum frequency resolution
and integration time ($\Delta\nu=\frac{1}{\Delta t}$) there is an effective
limit to the frequency resolution that can be obtained without correcting 
the received signal for this effect. 

There are other parameters of the signal that are unknown, for example: At what
frequency will it be transmitted?  What is the bandwidth of the 
signal?  Will the signal be pulsed, if so at what period?  Fully investigating 
a wide range of these parameters requires innovative instruments and 
enormous computing power.  Another possibility that, until now, has not 
been included in SETI searches is
that rather than directing large amounts of power into a narrow frequency band, 
an extraterrestrial intelligence might direct
large amounts of power into a narrow time window by sending a short duration
wide-band pulse.  A wide band signal has the advantage of reducing the 
importance of the choice of observation frequency, and the unavoidable
interstellar dispersion of a broad-band pulse reduces the confusion between
terrestrial and extraterrestrial pulses.  
In an ionized medium, high frequencies propagate slightly faster than low frequencies at radio frequencies.  Therefore RF pulses will be dispersed in frequency during their travel through the interstellar medium.  The dispersion across a bandwidth
$\Delta \nu$ is given by 
\begin{equation}
\delta t = (8.3 \us) \frac{\Delta\nu (\MHz)}{\nu^{3} (\GHz)} DM 
\label{DM_eq}
\end{equation}
where dispersion measure $DM$ is defined as $\int_0^L n_e dl$ and
is usually quoted in units of cm$^{-3}$ pc.

It is possible that a civilization intentionally creating a beacon
for extraterrestrial astronomers would choose to create ``pulses'' which have
a negative DM.  Natural dispersion always causes higher 
frequency components to arrive first.  A signal in which the low frequencies
arrive first would stand out as obviously artificial.  As a check on this,
as well as to establish our background noise limit, we examine both
positive and negative dispersion cases.
Unfortunately, sensitive detection of broad-band
pulses at an unknown dispersion measure also requires enormous computing power.

We are currently conducting two searches for these dispersed radio pulses.  Astropulse at Arecibo Observatory and Fly's Eye at the Allen Telescope Array (ATA).  The Allen Telescope Array is a joint project of the SETI Institute and the University of California, Berkeley \cite{deboer}.
\subsection{Pulses from Extraterrestrial Intelligence}

Astropulse and Fly's Eye consider the possibility that extraterrestrials communicate using high intensity and wideband but short timescale pulses.  It turns out that both the conventional (narrowband) method of data transfer assumed by SETI@home and the wideband, short timescale method assumed by Astropulse and Fly's Eye require the same amount of energy to send a message.

In each case, the energy required to send one bit of information is proportional to the energy (per area) required to send the minimum detectable signal.  And this energy is the same for the two methods, so energy considerations cannot rule out short timescale pulses as a medium for extraterrestrials' communication.

 SETI@home's sensitivity in searching for narrowband signals is given by:

\begin{equation}
\frac{\alpha \sigma T_{\textrm{sys}}}{G \sqrt{N_{\textrm{pol}} B t_{\textrm{int}}}}
\end{equation}

where $\alpha = 2$ is a loss from $1$-bitting the data twice, $\sigma = 24$ is the threshold above noise, $T_{\textrm{sys}} = 28$ K is the system temperature, $G = 10 \textrm{ K Jy}^{-1}$ is the gain, and $N_{\textrm{pol}} = 1$ is the number of polarizations.

SETI@home searches for narrowband signals with a bandwidth of $0.075$ Hz, and its longest integration time is ($0.075$ Hz)$^{-1} = 13.4$ s.  To compute the energy (per area) required to send one bit, multiply by $B t_{\textrm{int}} = 1$ to get $134 \cdot 10^{-26}$ J m$^{-2}$.

On the other hand, Astropulse and Fly's Eye are searching for short pulses, so the sensitivity of these experiments will be given in Jansky microseconds, by a slightly different formula:

\begin{equation}
\frac{\alpha \sigma T_{\textrm{sys}} \sqrt{t_{\textrm{int}}}}{G \sqrt{N_{\textrm{pol}} B}}
\end{equation}

where the variables have the meanings as before, except that now $t_{\textrm{int}}$ is the timescale of the pulse and B  the entire bandwidth.  

For Astropulse, $t_{\textrm{int}}$ is $0.4$ $\mu s$ and $B = 2.5$ MHz.  Also, $\alpha = 1.4$ because Astropulse $1$-bits the data only once, $N_{\mathrm{pol}} = 2$, and $\sigma = 21.5$.  The resulting sensitivity is $24 \textrm{ Jy } \mu$s.  To compute the energy per area required to send one bit, multiply by $B = 2.5 \cdot 10^6$ Hz to get $60 \cdot 10^{-26}$ J m$^{-2}$. 

For both narrow frequency and broadband pulse cases, the expression for the minimum detectable energy has the form

\begin{equation}
\frac{\alpha \sigma T_{\textrm{sys}} \sqrt{B t_{\textrm{int}}}}{G \sqrt{N_{\textrm{pol}}}}
\end{equation}

with $\sqrt{B t_{\textrm{int}}} = 1$, so similar answers are expected.

Because these energies are comparable, we have a chance to detect ETI communications in this new regime.  

\subsection{Pulses from Evaporating Black Holes}
Another intriguing possible source of short duration pulses comes from a suggestion by Martin Rees in 1977 \citep{rees} that primordial black holes, evaporating via the Hawking Process, could emit a large electromagnetic signature.  
According to Hawking \citep{hawk}, a black hole of mass $M$ emits radiation like a blackbody with a temperature $T_{BH}$ given by:
\begin{equation}
T_{BH}  = \frac{{\hbar c^3 }}{{8\pi kG_{grav}M}} = 10^{ - 6} \left( {\frac{{M_ \odot  }}{M}} \right)K
\end{equation}
This radiation emanates from the black hole event horizon and comes completely from the black hole's mass.  For a non-accreting black hole, Stefan-Boltzmann yields a lifetime of:
\begin{equation}
\tau _{BH}  = 10^{10} years\left( {\frac{M}{{10^{12} kg}}} \right)^3 
\end{equation}
For stellar mass black holes, this theory predicts lifetimes of order 10$^{34}$ years, much too long to ever expect to observe.  However, some cosmologies predict the creation of numerous small (M $\sim$ 10$^{12}$g) primordial black holes in the early universe, which according to theory could be evaporating now \citep{hawk2}.  

The specific mechanism by which the evaporating black hole produces a strong radio pulse has not been fully elucidated, but in short, it is thought that the process is similar to the EMP that accompanies supernova explosions.  In such a process, a highly conductive plasma fireball expanding into an ambient magnetic field can exclude the field and create an electromagnetic pulse.  For typical values of the interstellar magnetic field, this pulse would be peaked near 1GHz \citep{bla77}.

An observation of these pulses would not only provide a significant confirmation of Hawking radiation, but would also give strong evidence of the existence of primordial black holes.


\section{Fly's Eye: Searching for Bright Pulses with the ATA}


The Allen Telescope Array has several advantages over other telescopes worldwide for performing transient searches, particularly when the search is for bright pulses. The ATA has 42 independently-steerable dishes, each 6m in diameter.  The beam size for individual ATA dishes is considerably larger than that for most other telescopes, such as VLA, NRAO Green Bank, Parkes, Arecibo, Westerbork and Effelsberg.  This means that the ATA can instantaneously observe a far larger portion of the sky than is possible with other telescopes. Conversely, when using the ATA dishes independently, the sensitivity of the ATA is far lower than that of other telescopes.\\

The Fly's Eye instrument was purpose built to search for bright radio pulses of millisecond duration at the ATA.  The instrument consists of 44 independent spectrometers using 11 CASPER IBOBs. Each spectrometer processes a bandwidth of 210MHz, and produces a 128-channel power spectrum at a rate of 1600Hz (i.e. 1600 spectra are outputted by each spectrometer per second). Therefore each spectrum represents time domain data of length 1/1600Hz=0.000625s=0.625ms, and hence pulses as short as 0.625ms can be resolved\footnote{Pulses of duration $<$0.625ms can also be detected provided that they are sufficiently bright, but their length cannot be determined with a precision greater than the single spectrum length.}.

\begin{figure}[htp]
	\centering
		\includegraphics[width=0.9\linewidth]{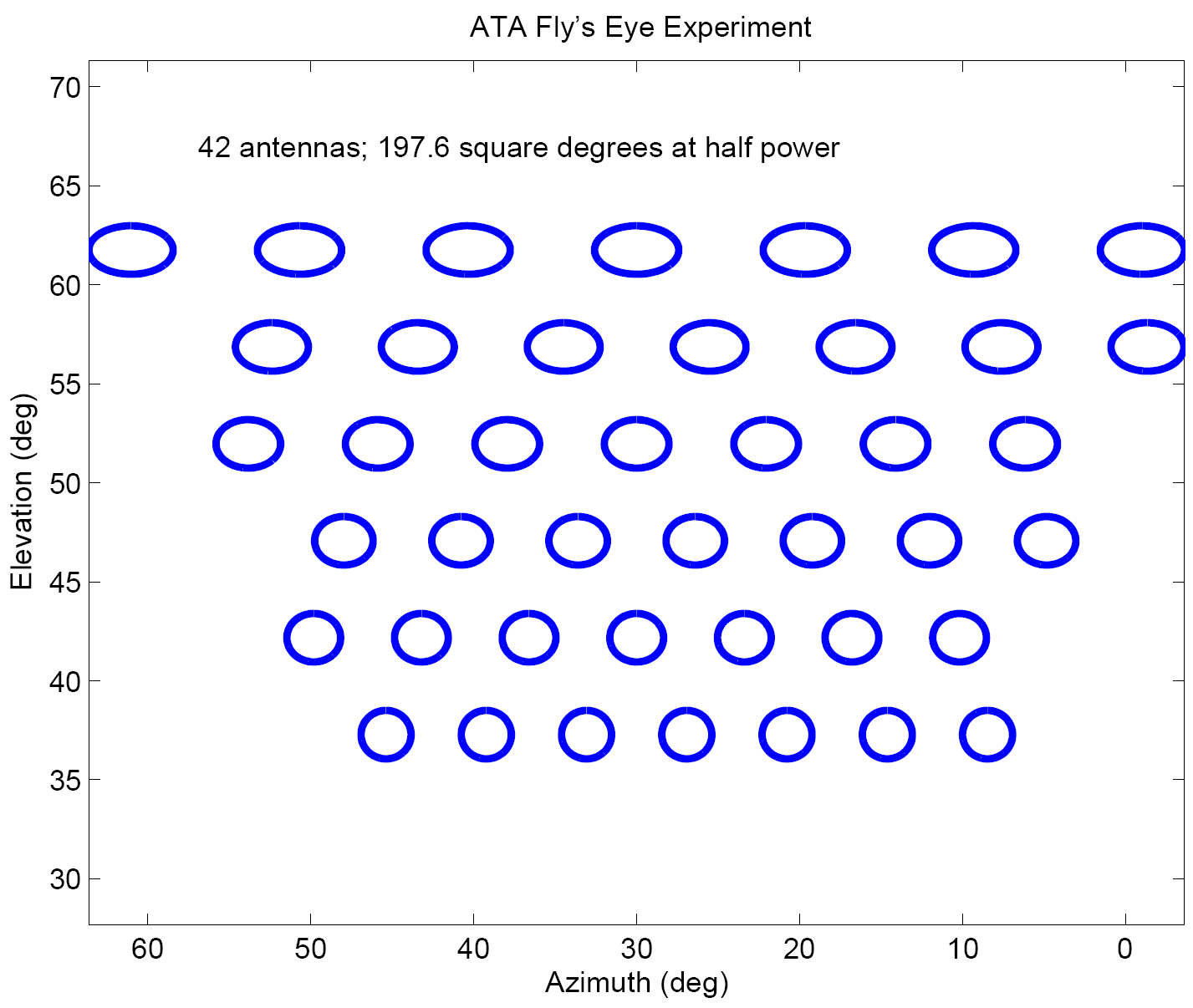}
		\caption[A hexagonal packing beam pattern for 42 antennas at the ATA.]{From \cite{BowerFlysEyeProposal08}. The beam pattern of the 42 beams at ATA, with the diameters equal to the half-power width. This hexagonal packing is pointing north. A south pointing results in poor interference properties, due to the highly populated areas south of the ATA.}
	\label{fig:FlysEyeNorthernPointingBeams}
\end{figure}

\begin{figure}[htp]
	\centering
		\includegraphics[width=0.9\linewidth]{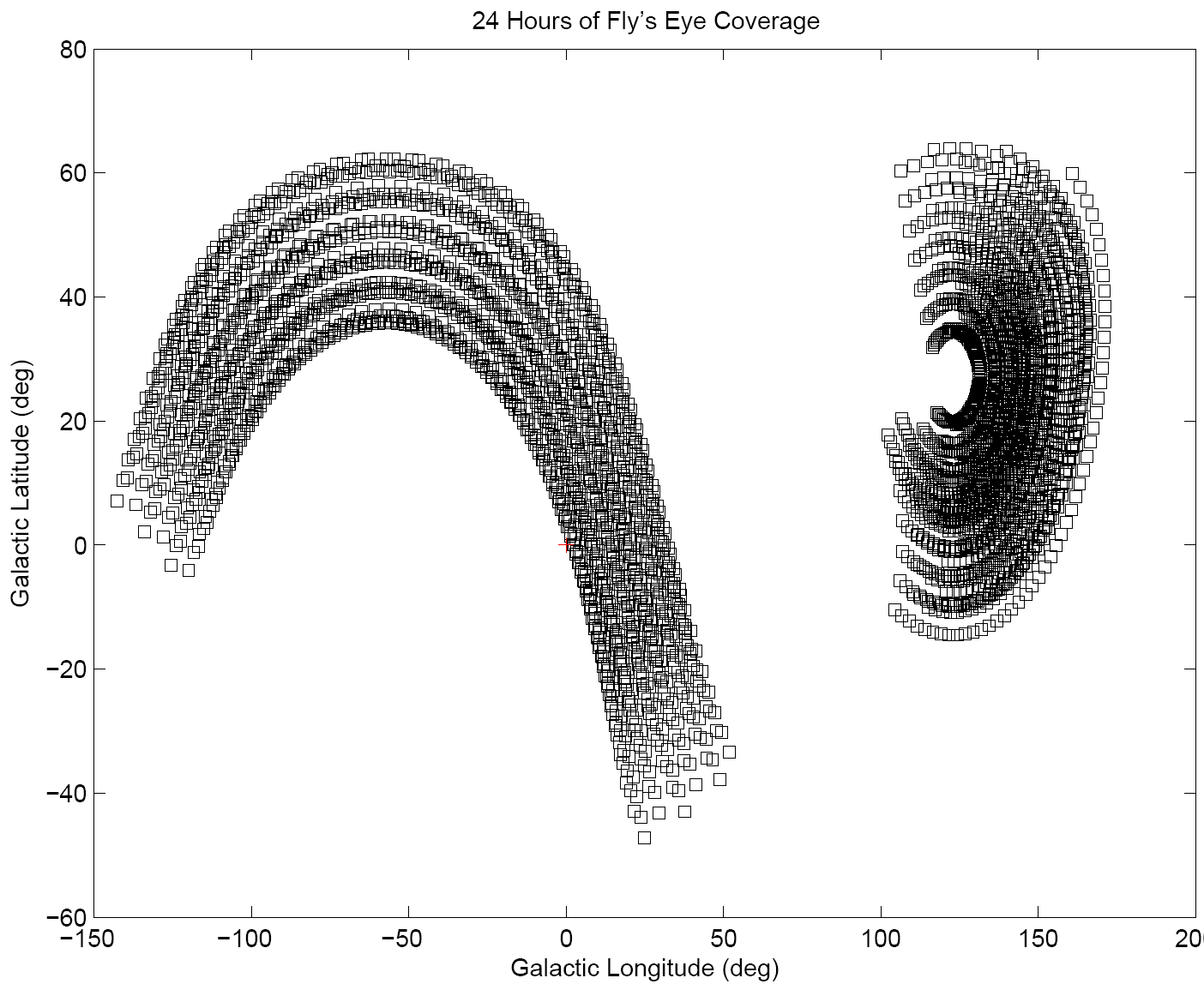}
		\caption[Sky coverage over 24 hours for 42 antennas at the ATA.]{From \cite{BowerFlysEyeProposal08}. The sky coverage of the ATA for an observing period of 24 hours. Both the coverages for southern and northern pointings (corresponding to the respective contiguous regions) are shown.}
	\label{fig:FlysEyeSkyCoverage}
\end{figure}

We have to-date performed roughly 400 hours of observing with the Fly's Eye.  Figures \ref{fig:FlysEyeNorthernPointingBeams} and \ref{fig:FlysEyeSkyCoverage} show the beam pattern and sky coverage using all 42 antennas.

\subsection{System Architecture}

\begin{figure}[htp]
	\centering
		\includegraphics[width=0.9\linewidth]{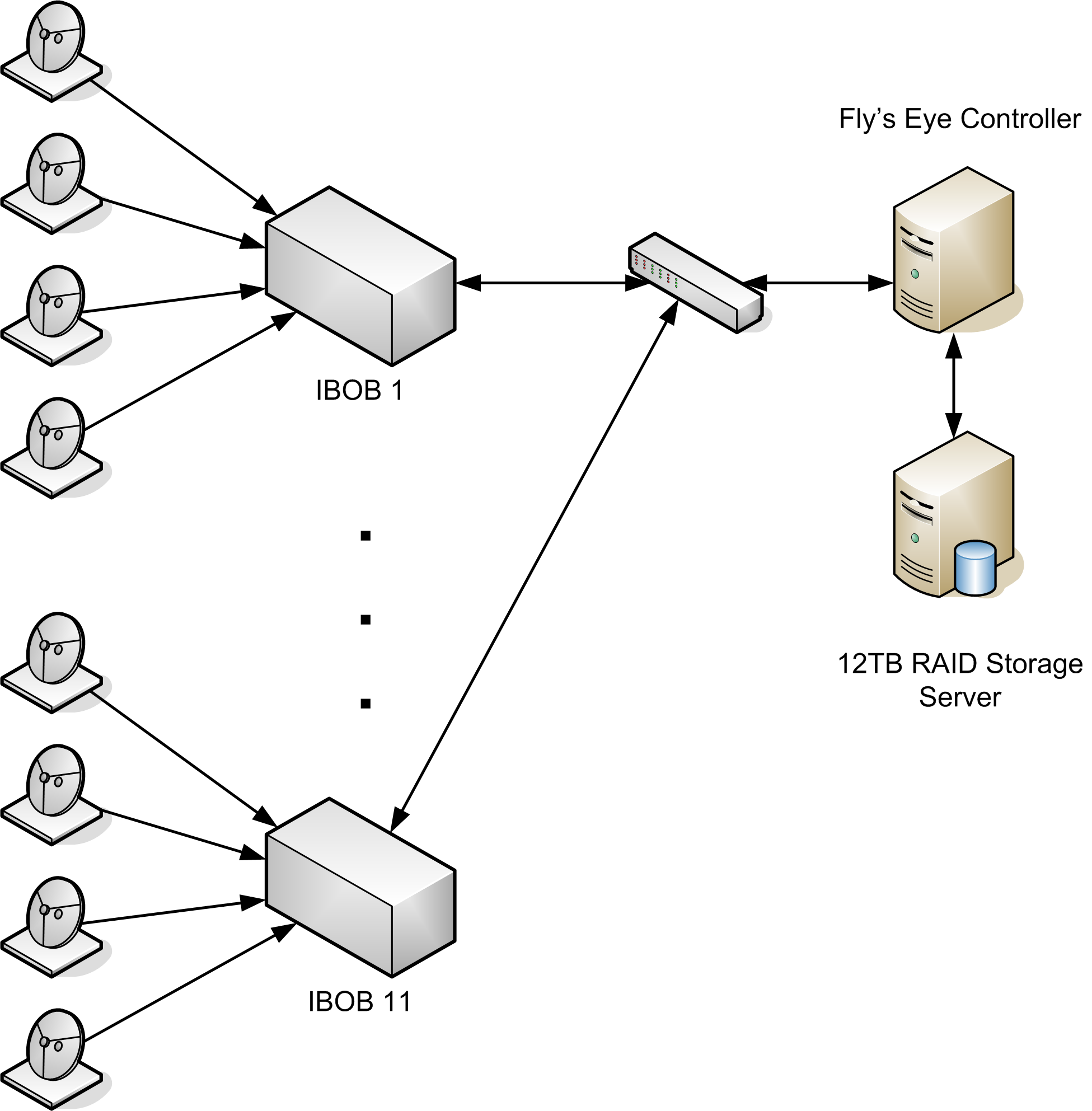}
		\caption[Fly's Eye System Architecture.]{Fly's Eye System Architecture. A selection of 44 analogue signals from 42 dual polarization antennas are connected to 44 independent spectrometers implemented in 11 IBOBs.}
	\label{fig:FlysEyeSystemArchitecture}
\end{figure}

The overall architecture of the Fly's Eye system is shown in Figure \ref{fig:FlysEyeSystemArchitecture}. Each IBOB can digitize four analogue signals, and 11 IBOBs are provided so that 44 signals can be processed. The ATA has 42 antennas, each with two polarization outputs. A selection\footnote{The selection is made with consideration for the goal of maximising field-of-view -- in practice we selected at least one polarization signal from every functioning antenna.} of 44 of the available 84 signals is made, and these are connected to the 44 iADC inputs. The IBOBs are connected to a control computer and a storage computer via a standard Ethernet switch.

\begin{figure}[htp]
	\centering
		\includegraphics[width=0.9\linewidth]{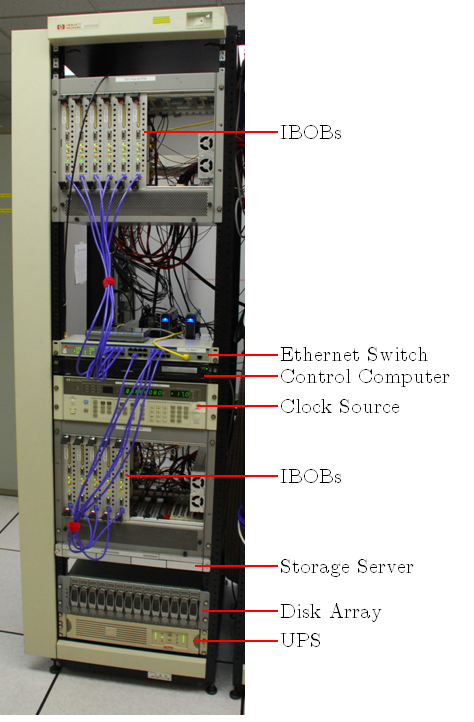}
		\caption[Fly's Eye Rack at the ATA.]{Fly's Eye Rack at the ATA. Two 6U CompactPCI crates (top and bottom) house the 11 IBOBs. The switch, data recorder computer and storage server are all visible.}
	\label{fig:FlysEyeRack}
\end{figure}

\subsection{Fly's Eye Offline Processing}

The analysis required for the Fly's Eye experiment is, in principle, fairly simple -- we wish to search over a wide range of dispersion measures to find large individual pulses. Specifically our processing requires that all the data be dedispersed with dispersion measures ranging from 50 cm$^{-3}$ pc to 2000 cm$^{-3}$ pc. At each dispersion measure the data needs to be searched for `bright' pulses. 

The processing chain is in practice significantly more complicated than this description suggests. Processing is performed on compute clusters, with input data formatted, divided and assigned to worker nodes for processing. In the worker node flow, the data is equalized, RFI rejection is performed, and finally a pulse search is performed through the range of dispersion measures. The results are written to a database where they can be subsequently queried. The key feature of the results is a table that lists, in order of decreasing significance, the pulses that were found and the dispersion measures they were located at. \\

Average power equalization is performed on the frequency spectrum equalized values $P_i'(t)$. We compute the average power over all frequency channels for a single integration (time sample $t$). The power average is defined as $\overline{P'(t)} = \frac{1}{N} \sum_{i=0}^{N-1} P_i'(t)$. $N$ is the number of channels (for Fly's Eye this is always 128). The motivation for why it is possible to normalize the power is that we expect pulses to be dispersed over many time samples, so this procedure should not remove extraterrestrial pulses. \\

Our strategy for mitigating constant narrowband RFI is simply to identify the channels that are affected, and to exclude them from further processing. This channel rejection is typically performed manually by looking at a set of spectra and identifying obviously infected channels, which are then automatically excluded in subsequent processing runs.\\

Intermittent RFI is often quite difficult to automatically distinguish from genuine astronomical pulses, and we followed a conservative approach to try to ensure that we do not accidentally excise dispersed pulses. Our statistic for intermittent RFI is the variance of a single channel over a 10 minute data chunk, $\sigma_i^2 = \left( \frac{1}{T_0} \sum_{t=0}^{T_0-1} \left(P_i''(t)\right)^2 \right) - \left( \frac{1}{T_0} \sum_{t=0}^{T_0-1} \left(P_i''(t)\right) \right)^2$. Curve-fitting determines a $\sigma_i^2$ outside which it is likely that channel $i$ contains time-varying RFI. Future reprocessing will likely use a more robust method, such as that based on a kurtosis estimator \cite{Nita07}. \\

Our final RFI mitigation technique is manual -- in our results it is easy to see high-$\sigma$ hits that are a result of RFI: these hits appear as simultaneous detections at many dispersion measures. \\

\subsection{Detection of Giant Pulses from the Crab Nebula}

A suitable test of transient detection capability is to observe the Crab pulsar and attempt to detect giant pulses from it. \\

\begin{figure}[htp]
	\centering
		\includegraphics[width=0.9\linewidth]{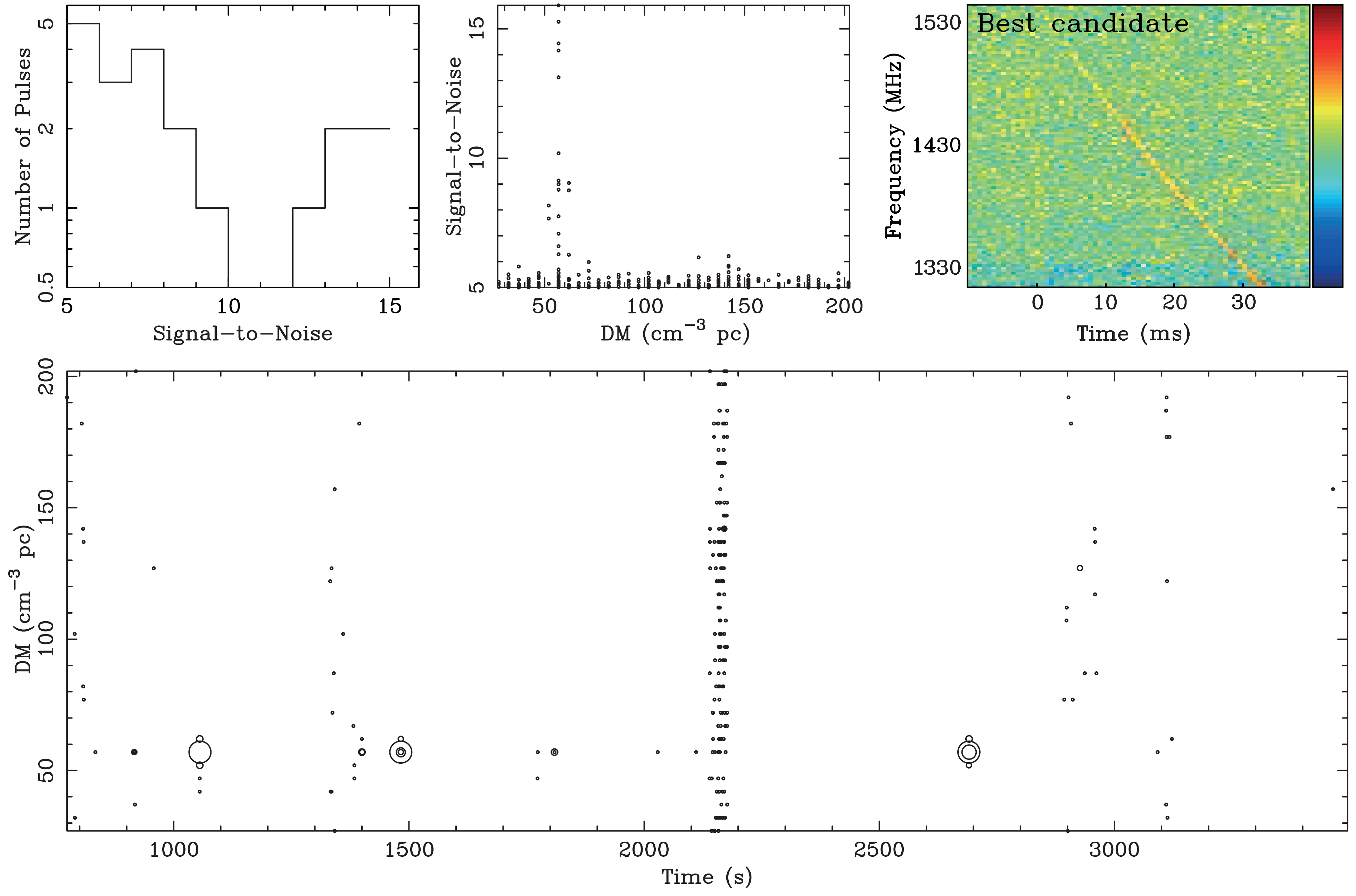}
		\caption[Crab Pulsar Observation Diagnostics.]{Diagnostics on data taken from the Crab pulsar in a 60-minute observation conducted on 22 December 2007. The data was dedispersed using dispersion measures ranging from 0 to 200 cm$^{-3}$ pc. Top-left: single-pulse SNR histogram. Top-centre: noise appears at all DMs, but bright pulses (SNR$>7$) from the Crab correctly appear at DM $\approx 57$ cm$^{-3}$ pc. Top-right: inset of Figure \ref{fig:FlysEyeCrabGiantPulse}. Bottom: pulse detections plotted on the DM versus time plane. Higher SNR detections appear as larger circles. Three giant pulses from the Crab are clearly visible in this plot.} 
	\label{fig:FlysEyeCrabGiantPulseDiagnostics}
\end{figure}

Figure \ref{fig:FlysEyeCrabGiantPulseDiagnostics} shows a diagnostic plot generated from a one-hour Crab observation. The data is an incoherently summed set from the 35 best inputs. The diagnostic plot was generated after the raw data had been dedispersed using a range of dispersion measures from 5 to 200. The Crab pulsar has dispersion measure $\approx 57$ cm$^{-3}$ pc, so three giant pulses from the Crab pulsar can be easily identified in the lower plot. The giant pulses appear only at the expected dispersion measure, whereas wideband RFI appears across a wide range of dispersion measures. \\

\begin{figure}[htp]
	\centering
		\includegraphics[width=0.9\linewidth]{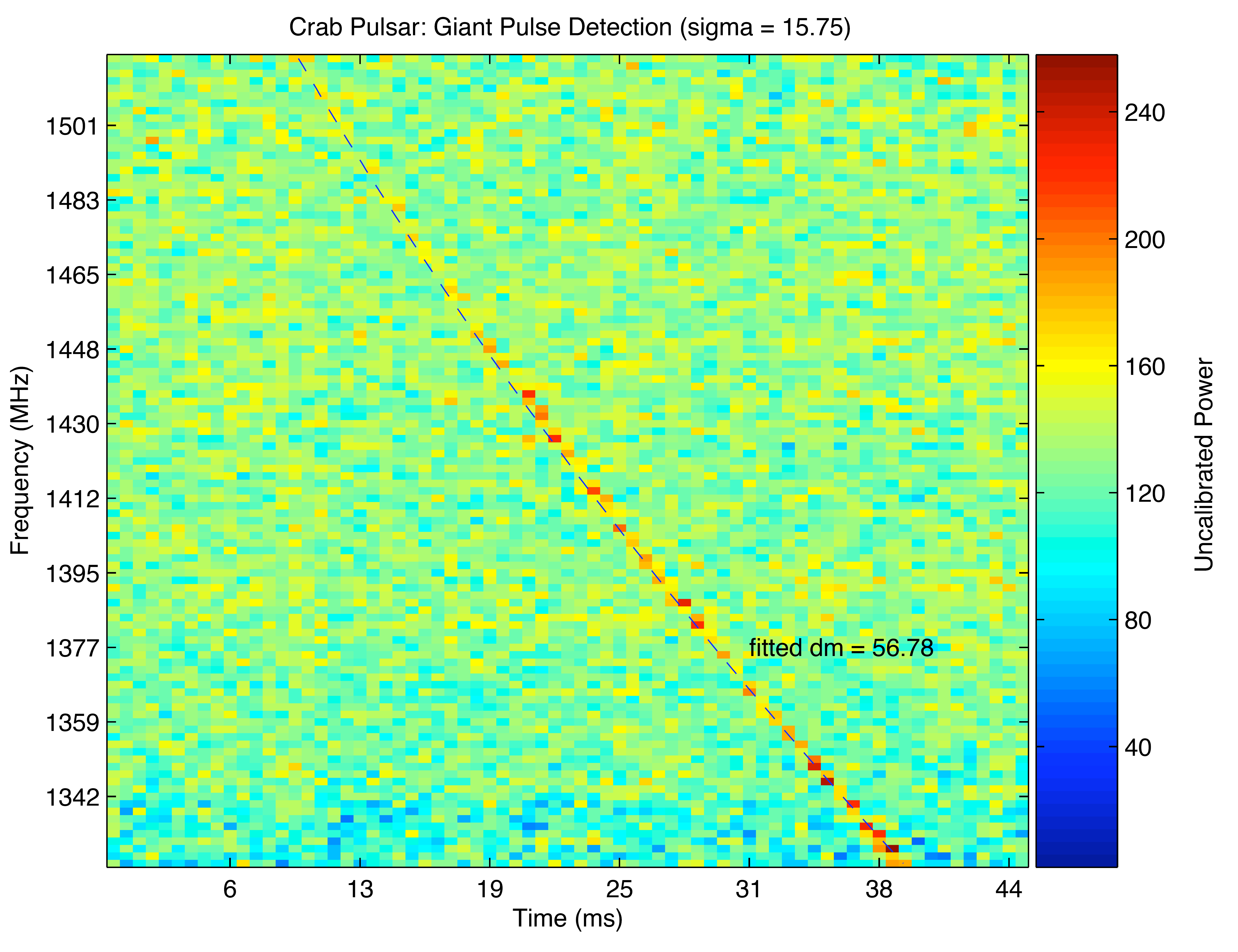}
		\caption[A Giant Pulse from the Pulsar in the Crab Nebula.]{A giant pulse from the pulsar in the Crab nebula. This pulse was detected in a 60-minute dataset taken on 22 December 2007. The dispersion of the pulse, correctly corresponding to DM = 56.78 cm$^{-3}$ pc, is clearly visible.}
	\label{fig:FlysEyeCrabGiantPulse}
\end{figure}

A frequency vs. time plot of the raw (summed) data at the time when the brightest giant pulses was detected is shown in Figure \ref{fig:FlysEyeCrabGiantPulse}. The pulse is clearly visible in the data, and a fit to the dispersion measure shows that the pulse is, nearly without doubt, from the Crab (as opposed to RFI). \\

\section{Astropulse}

\subsection{Algorithm}

\label{alg_sec}

Astropulse uses a unique approach to the problem of searching for dispersed
pulses.  This problem is well known to pulsar astronomers, and many
solutions have been devised over the years.  The traditional method
of searching for pulses of unknown dispersion measure ($DM$) is 
through incoherent de-dispersion, which is basically a filter bank - 
for each channel, the power is measured and integrated, then 
appropriate delays are 
added to compensate for the dispersion.  This can be made very efficient
through the use of a binary tree type algorithm.  
We use a coherent de-dispersion
algorithm, which preserves the phase of the signal, thus increasing 
sensitivity.  It also allows us to have much better time resolution
(down to the band limit at $0.4 \us$) than incoherent methods,
whose theoretical maximum time resolution is determined by the 
individual channel bandwidths.  (In practice, the time resolution for
pulsar searches is typically $100 \us$ or greater.)
The problem with coherent de-dispersion is that 
it is very computationally intensive.  We are able to afford this by 
implementing the pulse search in a distributed computing environment.

The coherent de-dispersion process is basically a convolution.
The time domain data needs to be convolved with an appropriate chirp
function in order to remove the dispersion smearing effect.

\begin{displaymath}
f_{ch}(t) = \exp (2\pi i \nu(t) t) 
      = \exp (2\pi i \frac{t}{\delta t} \Delta\nu t)
\end{displaymath}

Here, $\delta t$ is the amount of time stretching caused by dispersion
over a bandwidth of $\Delta \nu$, which can be determined from
equation \ref{DM_eq}.  


The most efficient way to do this is through the use of FFT convolution.
Here is a brief description of the Astropulse detection algorithm:

\begin{enumerate}
\item  Take $13.4 \s$ time-domain data, perform FFT on 32k-sample chunks.
       Overlap chunks by 50\% to recover pulses which span a chunk
       boundary.
       Save results on disk for later use.
\item  For each $DM$,
\begin{enumerate}
  \item  Multiply frequency domain data by correct chirp.
  \item  Inverse FFT to go back to the time domain.
  \item  Threshold the (de-dispersed) time data, recording strong pulses.
  \item  Co-add adjacent bins, and threshold to look for broad pulses.
  \item  Search repeating pulses by using a folding or harmonic search algorithm.
\end{enumerate}
\item  Repeat for next $DM$.
\item  When $DM$ range is covered, start at step 1 for next $13.4 \s$ of data.
\end{enumerate}
  
The optimal step size for $DM$ is to step by the amount of dispersion
which would cause 1 extra time sample length of stretching.  Using
the above expression for $\delta t = 0.4 \us$, this gives us a $DM$ 
resolution of $0.055 \dm$.  We want to cover $DM$s from $55 \dm$ to $830 \dm$, which
corresponds to 14,000 $DM$ steps at this resolution.  ($DM$s smaller than about $55$ are inaccessible to us, or to any search that employs one-bit sampling.)

The maximum $DM$ we wish to search determines the length of 
the FFTs.  
We want to look at $DM$ up to $830 \dm$, which will cover most
of the galaxy.  Using the same formula, this is a time stretch of
$\delta t = 5600 \us$, or about 14,000 samples.  Our FFT size should be
at least twice this big, so we will use 32k point FFTs. 

A FFT takes about $5N\log N$ floating point operations (FLOPs) to compute,
where $N$ is the number of points in the FFT.  We are doing $N=2^{15}$ point
FFTs, and we need to do one inverse transform for each $DM$, plus one original
forward transform.  In addition, there are another $N$ multiplies per
$DM$ for the chirp function, and a factor of 2 for the 50\% overlap.  
This makes our total number of operations 
for a 32k sample chunk of data:

\begin{displaymath}
N_{ops} = 2(N_{DM} + 1)N\log N + 2N_{DM}N = 1.8 \times 10^{10}
\end{displaymath}

To calculate how much computation we would need to analyze the data in real
time, divide the above number by $0.013 \s$ (length of 32k samples).  
We need another factor of 14 from multiple beams (7) and multiple polarizations (2).  Inefficiencies in memory allocation processes may contribute another factor of 5.  In all, 100 TeraFLOP/s or more may be required.  For comparison, an average desktop PC can compute
at a speed of about 1 GigaFLOP/s.

This analysis does not take into account the amount of computation required
for folding.  Astropulse folds on two different time scales, spending enough time on these to triple the overall computation requirement.
 
This computation would take far too long on one computer, or even on
a modest sized Beowulf cluster.  Our solution to the problem is to 
distribute the data publicly to volunteers who download a program
which will analyze it on their home computers.  This is set up as
a screen saver, which will turn on and start computing when their 
computers would
otherwise be idle.  \cite{korpela01}  This approach
has been remarkably successful for SETI@home, which has an average 
computation rate of about 40 TeraFLOP/s.

We are able to implement this though the BOINC software platform, also
developed by our group at Berkeley.  The package, the Berkeley
Open Infrastructure for Network Computing (BOINC), is a general
purpose distributed computing framework.  It takes care of all the
``management'' aspects inherent to a distributed computing project - 
keeping records of user accounts, distribution of data and collection
of results over the network, error checking via redundant processing,
sending out updates of the science code, etc.  Many lessons learned
over the course of running the original SETI@home project went
into BOINC, so that new groups wishing to start computing projects
don't have to ``re-learn'' these.

An interesting aspect of running a public distributed computing project
is that it has education and public outreach automatically built into it.
People invest their computer time into the project, and as a result become
interested in learning more about the science.  SETI@home has been enormously
successful in this regard - the website averages 1.25 million hits per day, and 
the project has spawned many internet discussion groups and independent 
websites.  It is also used as part of the science curriculum by thousands of
K-12 teachers nationwide.

\bibliographystyle{plain}

\end{document}